\documentclass[fleqn]{article}
\usepackage{amsmath}
\usepackage{graphicx}
\begin{document}

\title{\large  History-dependent synchronization in the Burridge-Knopoff model}   
 
\author{\normalsize
J.~Szkutnik,
B.~Kawecka-Magiera,
K.~Ku{\l}akowski$^*$\\
\normalsize \em
Faculty of Physics and Nuclear Techniques, AGH University of Science and
Technology\\
 \normalsize \em
al. Mickiewicza 30, 30059 Krak\'ow, Poland.\\
\normalsize \em
$ ^*$Corresponding author. E-mail: kulakowski@novell.ftj.agh.edu.pl
}

\maketitle
\thispagestyle{empty}

A three-blocks Burridge-Knopoff model is investigated. The dimensionless velocity-dependent
friction force $F(v)\propto (1+av)^{-1}$ is linearized around $a=0$. In this way,
the model is transformed into a six-dimensional mapping
$\mathbf{x}(t_{n})\to \mathbf{x}(t_{n+1})$, where $t_n$ are time moments when a
block starts to move or stops. Between these moments, the equations of motion
are integrable. For $a<0.1$, the motion is quasiperiodic or periodic,
depending on the initial conditions. For the periodic solution, we observe a
synchronization of the motion of the lateral blocks. For $a>0.1$, the motion
becomes chaotic. These results are true for the linearized mapping, linearized
numerical and non-linearized numerical solutions.\\
 

]
\section{Introduction}  

We consider the Burridge-Knopoff model (BKM) of three blocks \cite{vie}. This
is a simple chain of blocks connected with springs, with additional springs
which drive each block with a constant velocity $V$. The blocks move on a
surface with friction. The model belongs to a rich branch of many-body
dissipative systems: the train model \cite{vi}, the Feder and Feder model
\cite{ff} and the Olami-Feder-Christensen model \cite{ofc}. There are also
some connection from this branch to the Frenkel-Kontorova model \cite{braun}.
At the root of the whole Tree of many-body nonlinear systems there is the
fameous Fermi-Pasta-Ulam model  \cite{ford}. There is also a conservative
branch, with the Toda chain \cite{toda}, ding-dong \cite{dd} and ding-a-ling
\cite{dal} models and presumably some others as leaves.  

Most characteristical effect displayed by the dissipative branch of the Tree is the
stick-slip dynamics: the blocks move or stop at various time moments,
periodically for one block and perhaps chaotically for two or more blocks. Recent works
of Persson \cite{pr1,pr2} allow to treat BKM as a generic model of two
surfaces, one sliding on another with friction. On the other hand, the
stick-slip dynamics has been observed experimentally in nanoscopic systems
\cite{hen}. In particular, this kind of motion turned out to be relevant for
the behaviour of tips of the scanning probe microscopes.

The origin of the stick-slip dynamics has been often discussed in terms 
of the state and the rate dependent model \cite{rr,b1,b2}. There, the character
of the velocity dependence of the friction force, $F(v)$, is considered 
to be crucial. On the other hand, as it was demonstrated in \cite{vi}, 
the stability of the uniform motion depends only on the sign of the 
drivative $F'(v)$.

Usually, the discussion of all the above listed models concentrates on two topics: their
application to theory of earthquakes (which are at the origin of the problem
\cite{ehqs}) and the effects of self-organized criticality \cite{bak}. Both
topics overlap at the Richter-Gutenberg law, which is an example of a scaling
relation which has been observed out of laboratory \cite{rg}. Other effects
under discussion within the above models are chaos and synchronization
\cite{vie}. All these topics are often considered for large systems, despite
the fact that even small systems of this kind are not entirely known. The aim
of this work is to show some unexpected properties of the Burridge-Knopoff
system of three blocks. 

\begin{figure}[t]
\includegraphics[scale=.34]{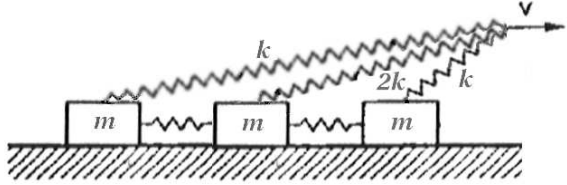}\\
\caption{The Burridge-Knopoff system considered in this work. Three equal
masses $m$ are drawn along a rough platform with constant velocity $V$.
 The masses are connected with springs of strength $k$. Similar springs 
 are used to attach them to the driving mechanism.}
\end{figure}
\begin{figure}[t]
\includegraphics[scale=.63]{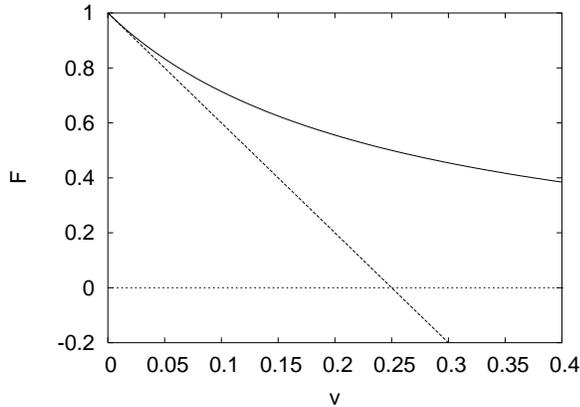}\\
\caption{Velocity dependent function of friction force $F(v)$.
Solid line for nonlinear form, dashed line for linearized form. 
Dotted line is a guide for eye.}
\end{figure}

In its original version \cite{vie} the BKM is nonlinear because of two reasons.
First is that the friction force $F$ acting on a block depends on its velocity
$v$ in a nonlinear way, i.e. $F=F(0)/(1+cv)$. Second is that the
equations of motion are nondifferentiable at time moments when a block starts
to move or stops. In our previous paper \cite{jskk} we have demonstrated that
for two blocks and small values of the parameter $c$, the first cause of
nonlinearity is not relevant. Namely, we substituted the nonlinear function
$(1+cv)^{-1}$ by its linear approximation $1-cv$. This step is justified for
small values of $c$, i. e. for strong spring constant $k$, 
large mass  $m$ and low friction force which only weakly 
depends on the velocity $v$. On the other hand, this step allows to integrate 
the equations of motion analytically. 
It is useful to introduce here a new variable $a=cF(0)/\sqrt{km}$.
It has been shown in Ref. \cite{jskk} that for $a<0.3$,
the results of the original and linearized versions are practically the same.

The goal of the present work is twofold. First observation is that the above
mentioned linearization does not change the results of BKM for three blocks,
when the parameter $a$ is less than about $0.1$. Second observation is that the
character of the motion can depend on the history of the system. Near $a=0.1$
we observe a transition from periodic or quasi-periodic behaviour for small $a$
to chaos for larger $a$. The character of the motion for $a<0.1$ does depend
on the initial conditions. Changing only the initial position of one of the
blocks, we switch from a quasi-periodic and non-synchronized motion to a
periodic evolution with the synchronization of two lateral blocks. This
switching from one kind of motion to another is the same for the linearized
motion and for the fully nonlinear formulation. In the latter case, the only
solution is numerical. Here we use the Runge-Kutta method of fourth order.

The paper is organized as follows. In the next section we describe the
linearization procedure. The outcoming equations are piecewisely linear and
they are solved analytically. The solution is valid except the time moments,
when the number of moving blocks varies. These time moments are found
numerically. In Section III we show the results obtained within the linearized
model and within its initial formulation - the latter is worked out
numerically. The last section is devoted to conclusions.

\section{The linearization}

The starting point for the linearization procedure is the set of three
equations of motion for three blocks

\begin{equation}  \label{eq_newt} 
\begin{split}
m\ddot{X}_1&=k(X_2-2X_1+Vt)-F(c\dot{X}_1)\\
m\ddot{X}_2&=k(X_1-4X_2+X_3+2Vt)-F(c\dot{X}_2)\\
m\ddot{X}_3&=k(X_2-2X_3+Vt)-F(c\dot{X}_3)\\
\end{split}
\end{equation}  
where $F(c\dot{X})=F(0)/(1+c\dot{X})$, $X_i$ is the position of $i$-th block,
$Vt$ is the position of the driving mechanism. The driving spring constants
are chosen to be $k$, $2k$ and $k$ for the first, second and third block
respectively, and the spring constants of the springs between the blocks are
all equal to $k$. Equation for $i$-th block is valid as long as it moves.
When the block stops, the condition of moving it again is that the total
spring force acting on the block must be greater than $F(0)$.\\
If according to \cite{vie} we change spatial and time units as follows: 
$\tau=\omega t$, $U_j=kX_j/F(0)$, $\omega^2=k/m$ we can
write the equations of motion in dimensionless form: 
\begin{equation}  \label{eq_tr} 
\begin{split}
\ddot{U}_1&=U_2-2U_1+v\tau-\Phi(a\dot{U}_1)\\
\ddot{U}_2&=U_1-4U_2+U_3+2v\tau-\Phi(a\dot{U}_2)\\
\ddot{U}_3&=U_2-2U_3+v\tau-\Phi(a\dot{U}_3)\\
\end{split}
\end{equation}
where $v=V\sqrt{km}/F(0)$ and $a=c F(0)/\sqrt{km}$.
The friction force has now the following shape: $\Phi(a\dot{U})=1/(1+a\dot{U})$.\\ 
For the linearized equations, the only modification is that the friction force
$\Phi(a\dot{U})=1-a\dot{U}$. This approximation is justified as long as the
constant $a$ is small. Additionally, we have to check whether the friction
force changes sign - this would be a harmful artifact. The profit is that the
equations of motion are analytically integrable: the functions have this
 shape:
\begin{equation}  \label{eq_shape} 
U_n=A \exp(\alpha\tau)+B \exp(\beta\tau)+C \exp(\gamma\tau)+D\tau+E
\end{equation}
We need to solve a new set of equations whenever one of the block starts 
to move or stops. This time moments are roots of function
(\ref{eq_shape}) and they are to be found numerically.
\begin{figure*}
\includegraphics[scale=0.9]{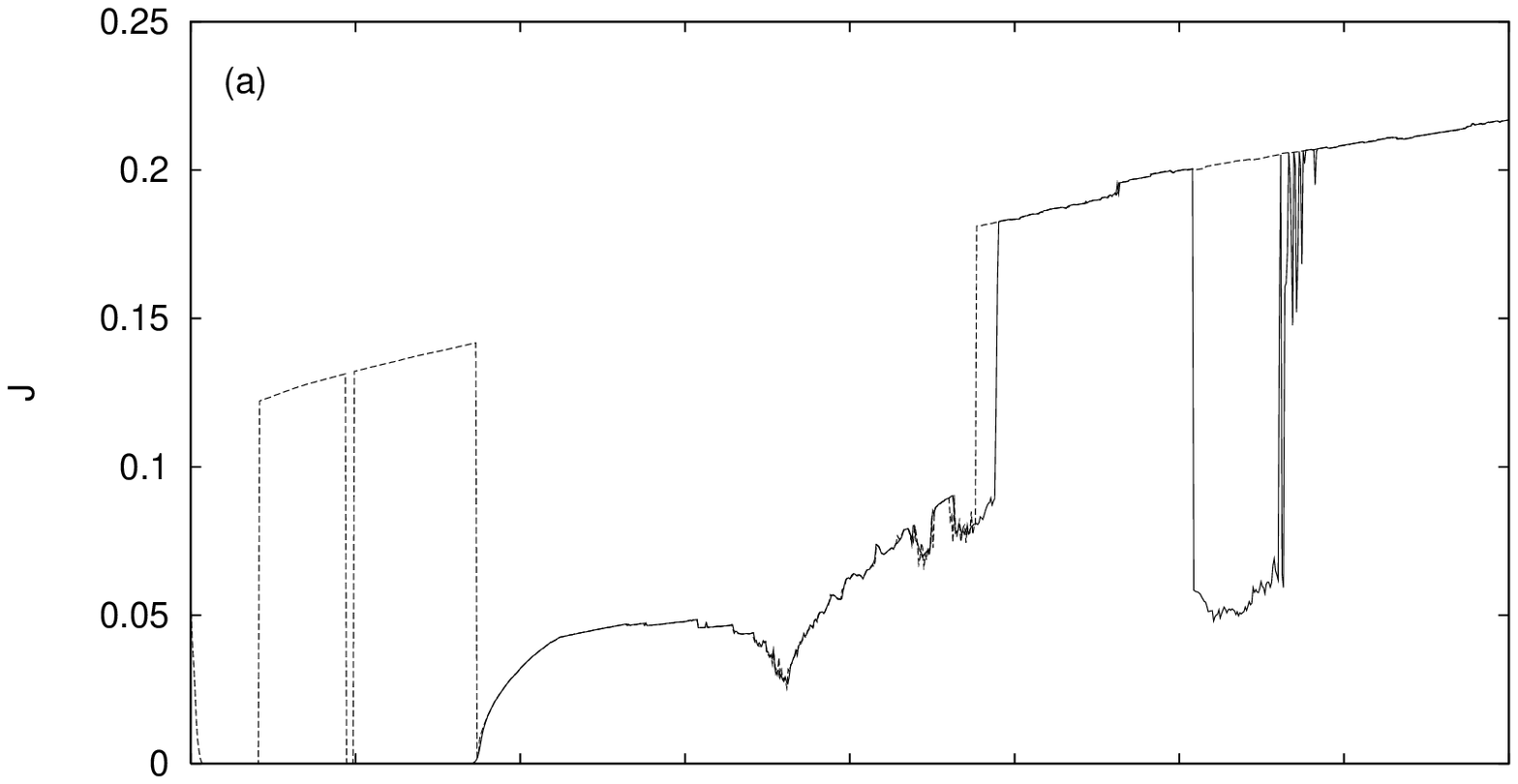}\\
\includegraphics[scale=0.9]{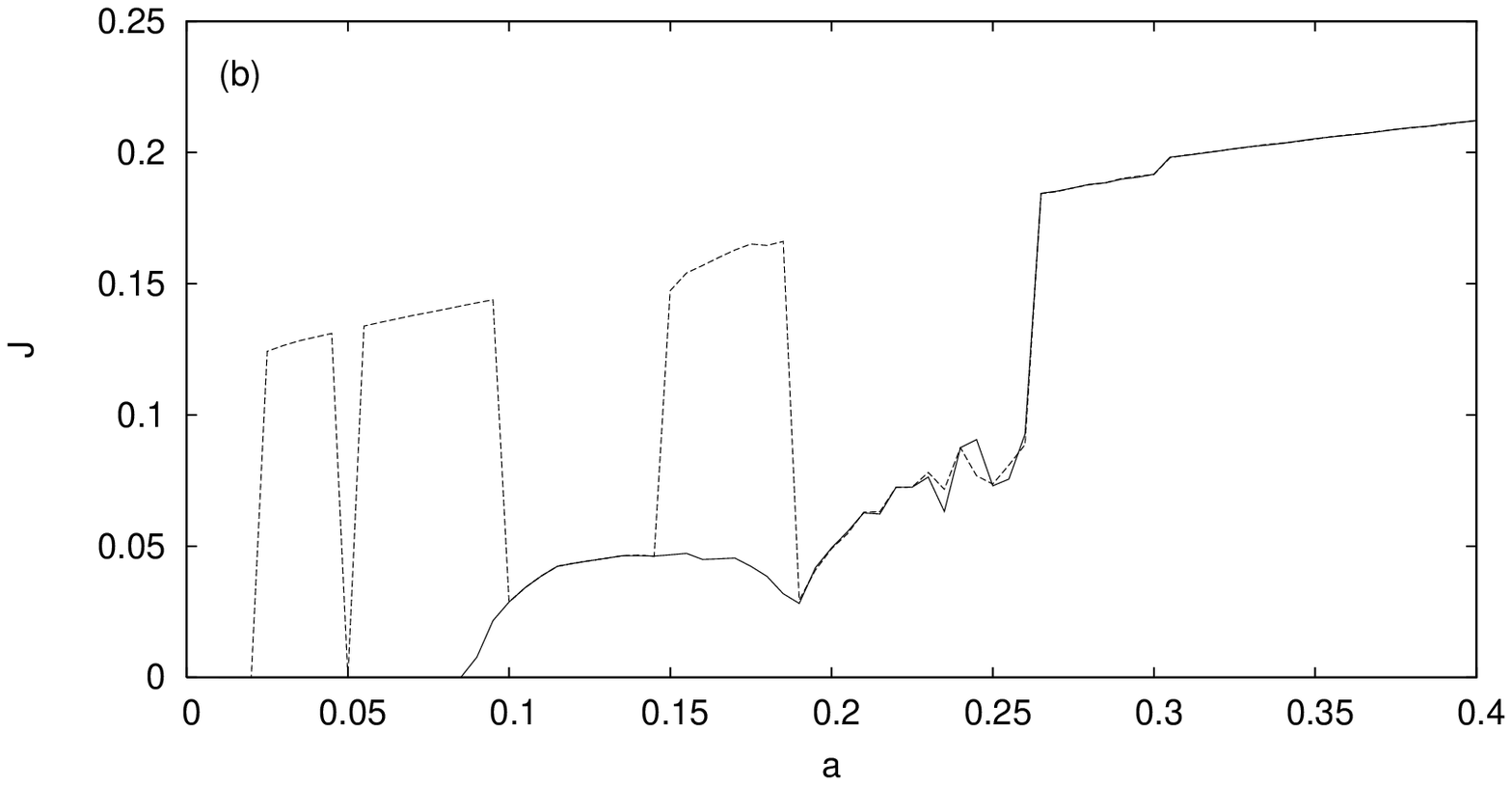}
\caption{The Euclidean distance $J$ in phase space beetween blocks 1 and 3 as function of
$a$. The solid line is for
$U_1(\tau=0)=+0.989$ in the initial state, and the dotted line - for $U_1(\tau=0)=-0.989$ in the initial state.
 (a) for linearized equation of motion (analytical solution) (b) for nonlinearized
equation of motion (numerical solution).}
\end{figure*}
\begin{figure*}
\includegraphics[scale=0.9]{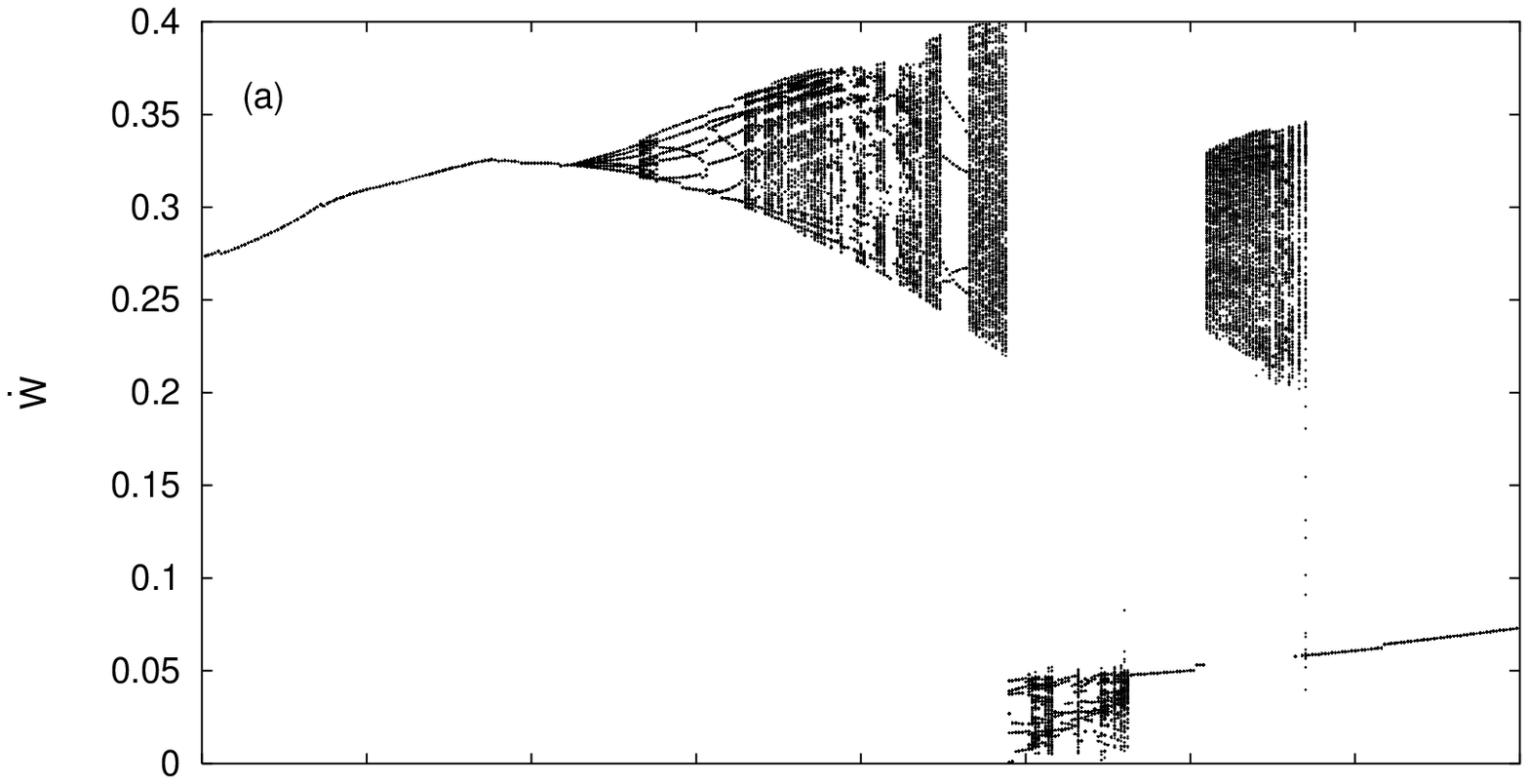}\\
\includegraphics[scale=0.9]{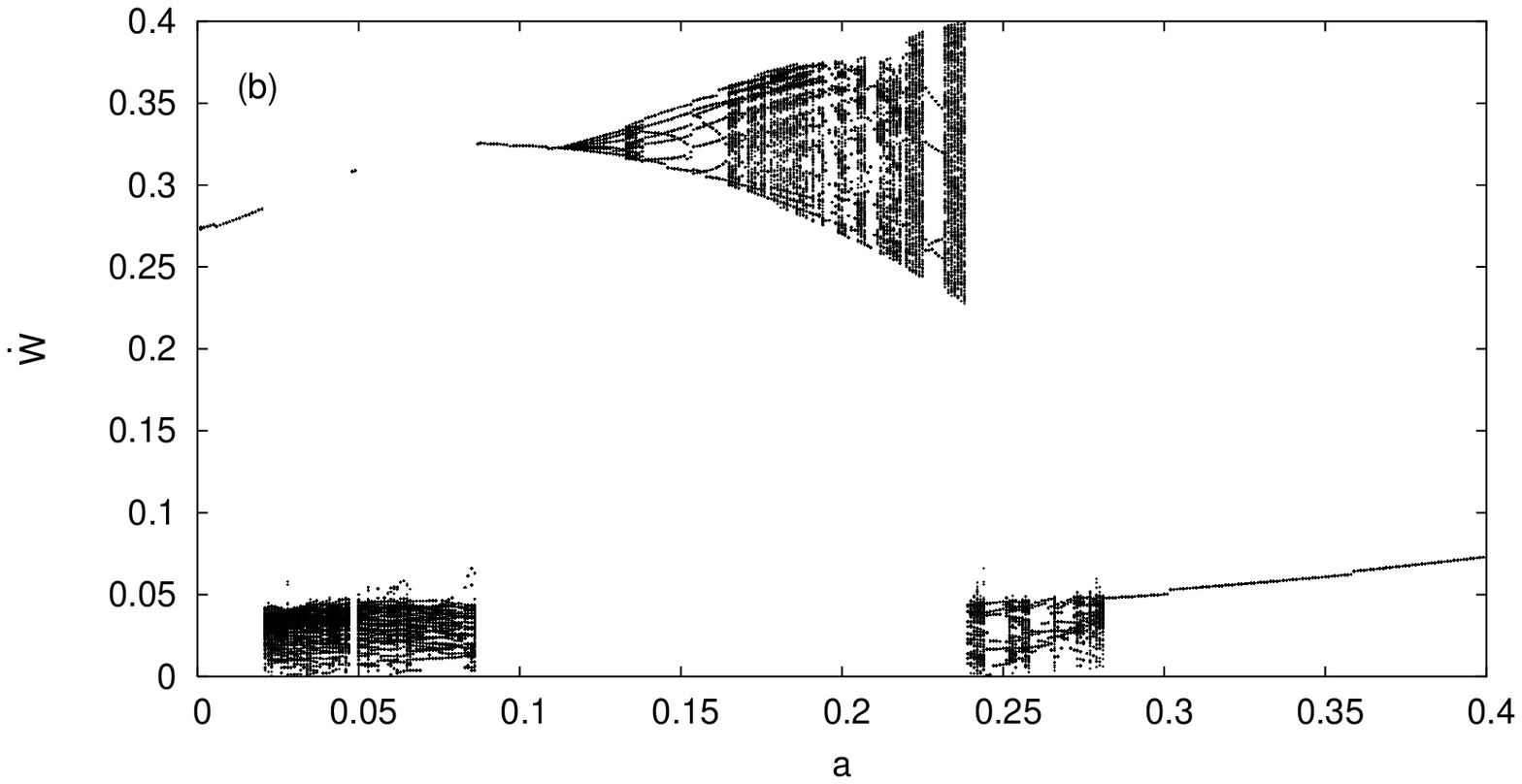}
\caption{The bifurcation diagram of $\dot{W}$
in the Poincar\'e section in which
$W=0$ (a) for $U_1(\tau=0)=-0.989$ in the initial state
(b) for $U_1(\tau=0)=+0.989$ in the initial
state.}
\end{figure*}
\section{Results} 
In Figs. 3a and 3b we show the time averages of the distance $J(\tau)$ in the
phase space between the first and third block. This distance is defined as 
\begin{equation}  \label{eq_j} 
J(\tau)=\sqrt{(\dot{U}_3(\tau)-\dot{U}_1(\tau))^2+(U_3(\tau)-U_1(\tau))^2}
\end{equation}
and its time average is performed over a time $\Delta \tau=7\times 10^3$ after a
transient of $3\times 10^3$. In the numerical calculations, each time unit is
$10^6$ time steps. In both analytical and numerical calculations, one
stick-slip cycle is about two or three time units. A comparison of the Figs. 3a
and 1b shows that there is a close correspondence between the results for the
initial nonlinear formulation and the linearized one. In both cases, the
synchronization appears for the parameter $a$ between $0.005$ and $0.08$. In both
cases, this synchronization does depend on the initial conditions in the same
way. When the initial position of the first block $U_1(\tau$=0)=-0.989, i.e. it
is close to the position for the stationary state, the motion is synchronized
and $<J(\tau)>=0$. (Here, $<...>$ is the time average.) This is so despite the fact
that the stationary motion is unstable. However, for $U_1(\tau=0)=+0.989$, the
motion is not synchronized and $<J(\tau)>$ is positive.

We note that the unstable character of the stationary motion can be proved in
the same way as it was done for two blocks by Vieira \cite{vi}. As long as the
friction force decreases with velocity, some eigenvalues of the Jacobian are
always positive.
\begin{figure}[t]
\includegraphics[scale=.6]{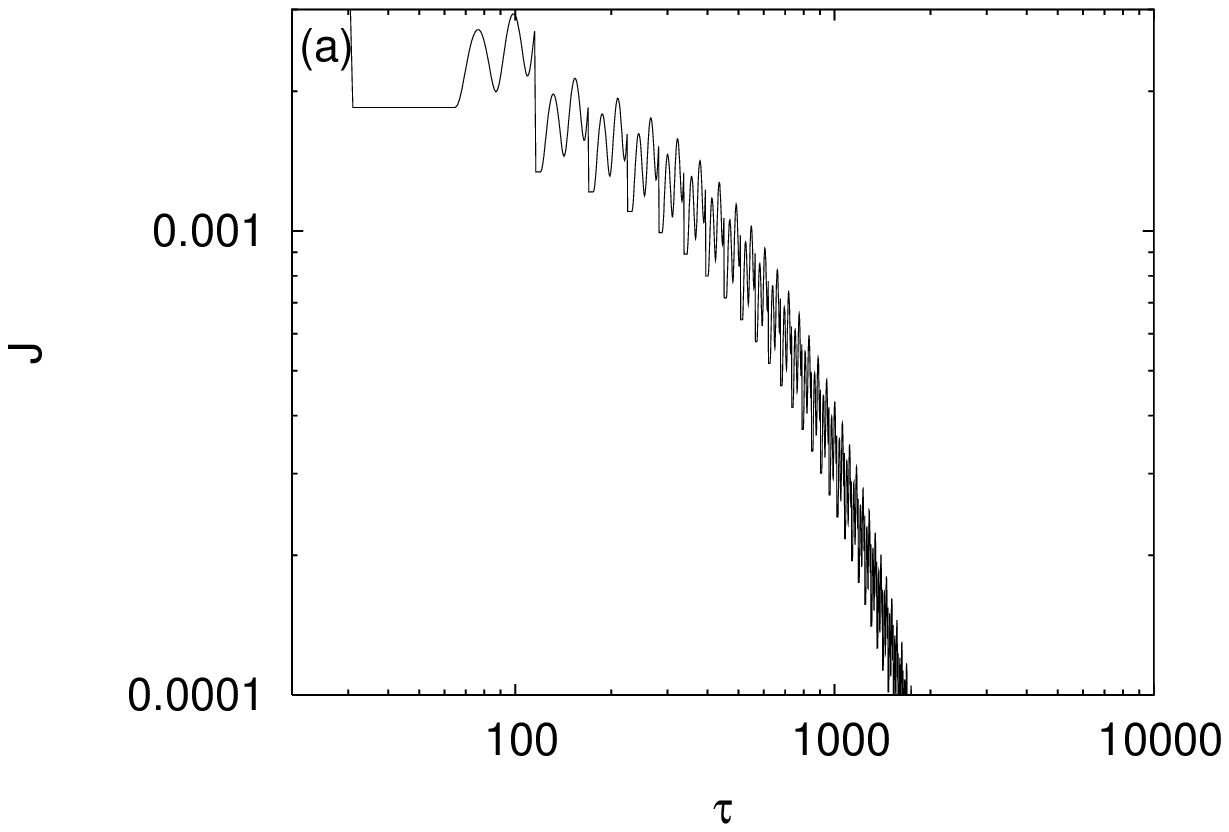}\\
\includegraphics[scale=.6]{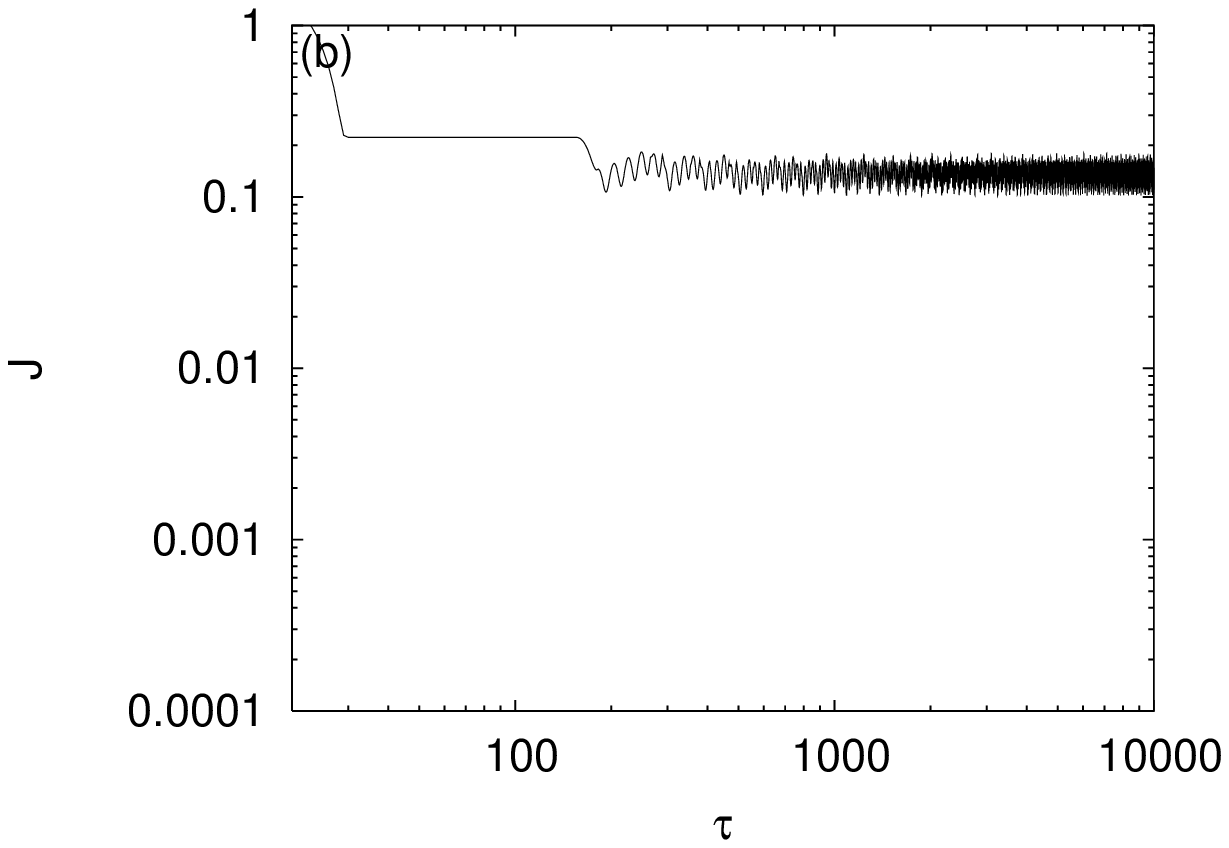}
\caption{The instantaneous distance in the phase space beetween the lateral blocks (a) for
initial position of the 1st block $U_1(\tau=0)=-0.989$, (b) for initial position of the 1st block
$U_1(\tau=0)=+0.989$.}
\end{figure}
The comparison of the Figs. 3a and 3b allows to believe that the linearization
procedure does not alter the results for $a<0.1$. Other results presented
in Figs. 4-7 are obtained for the linearized equations of motion and for the
parameter $a=0.08$. We compare the results for the two above values of the initial
position of the first block. The initial positions of the other blocks are
$U_2(\tau$=0)=-0.993 and $U_3(\tau$=0)=-0.991. The initial velocities are 
$\dot{U}_1(\tau$=0)=0.099, $\dot{U}_2(\tau$=0)=0.1012 and $\dot{U}_3(\tau$=0)=0.1023. The initial values of the velocities are
very close to the driving velocity $v$, which is set as $0.1$. 

To construct the bifurcation diagrams, we use the sum of positions of the blocks
with respect to the equilibrium positions: 
\begin{equation}  \label{eq_bif}
W=U_1-U^e_1+U_2-U^e_2+U_3-U^e_3
\end{equation}
We denote $U_1^e=v\tau-5(1-av)/6$, $U_2^e=v\tau-2(1-av)/3$ and $U_3^e=v\tau-5(1-av)/6$ the unstable point of equilibrium,
which is found by taking $\ddot{U}_n =0$ and $\dot{U}_n=v$ in Eq. (\ref{eq_tr}). The stability of this solution can be calculated
as in \cite{vi}. In Figs. 4a and 4b we show the bifurcation diagrams for the synchronized and
non-synchronized motion. Subsequent points mean the Poincar\'e surface section of
 $\dot{W}$ at $W=0$. In the former case (Fig.4a, $U_1(\tau$=0)=-0.989) and below $a=0.1$ we see a
continuous line which means one fixed point of the mapping. This means that
the whole motion is periodic. In the latter case (Fig.4b, $U_1(\tau$=0)=+0.989),
we see that below $a=0.1$ the above line disappears. Instead, we observe a
cloud of points for lower values of the velocity. This means the
quasiperiodic or the chaotic motion. To check it, we have calculated the
maximal Lyapunov exponent and we have found it to be close to zero.

In Figs. 5a and 5b we show the time dependence of the distance $J(\tau)$
between the lateral blocks in the two above cases. In the synchronized case
(Fig.5a) $J(\tau)$ tends to zero as soon as for $\tau=4000$, whereas for the
nonsynchronized case $J(\tau)$ displays stationary fluctuations around $0.14$.
Here again, the whole difference is the initial position of the first block.
\begin{figure}[t]
\includegraphics[scale=.6]{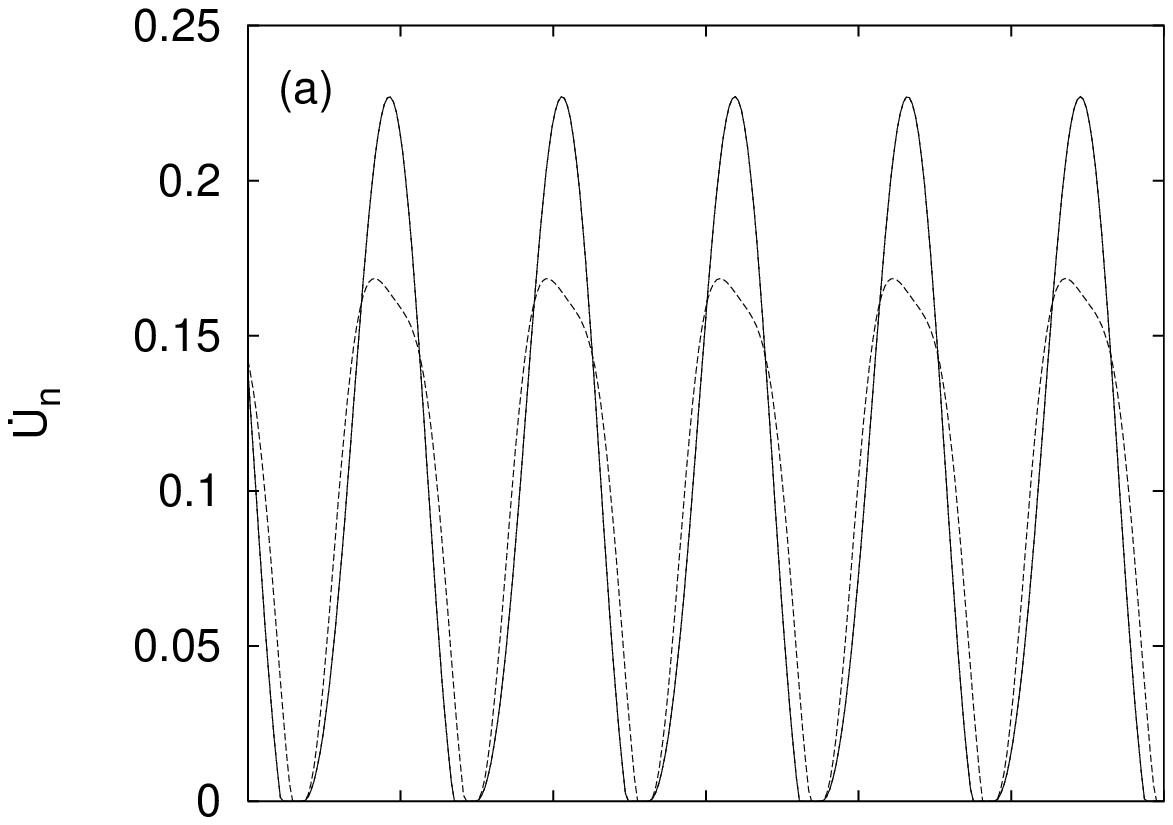}\\
\includegraphics[scale=.6]{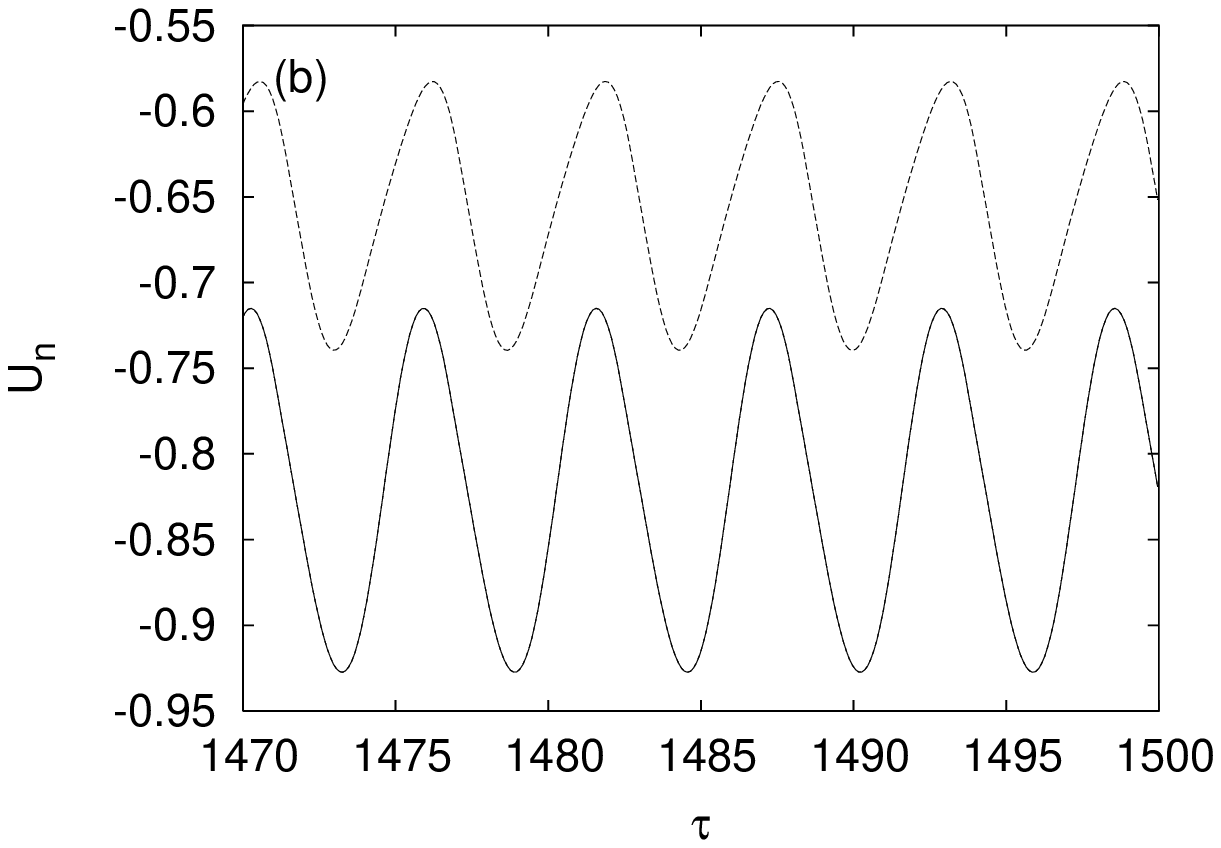}
\caption{(a) Velocities of blocks as a function of time $\tau$. (b) Positions of blocks as a function of time t.
 The initial position of the first block $U_1(\tau=0)=-0.989$.
  Solid line for blocks 1 and 3, dashed line for block 2.}
 \end{figure}
Finally, in Figs. 6 and 7 we show the character of the synchronized and
non-synchronized motion. In the former case (Fig. 6), the positions and
the velocities of the block are periodic functions of time. In the latter case
(Fig. 7), the motion is aperiodic. Moreover, the velocities of first
and third block seem to be in some kind of antiphase: when one increases,
another decreases. Both oscillate with the frequency more or less two times
smaller, than the middle block.

As we have seen in Figs. 6a and 7a, the stick-slip effect is present 
in both cases of periodic and aperiodic motion. The plots of velocities vs. time 
touch the value zero for each of the three blocks. It seems that some 
sticks are unavoidable. This conclusion is supported by the fact, that the
analytical solution for the linearized mapping contains the term 
$\exp[-F'(v)t]$, where $F(v)$ is the velocity dependence of the friction 
force. Therefore, if only the derivative $F'(v)$ is negative, the amplitude
of the velocity oscillations increases with time and $v$ must reach the 
zero value.

\begin{figure}[t]
\includegraphics[scale=.6]{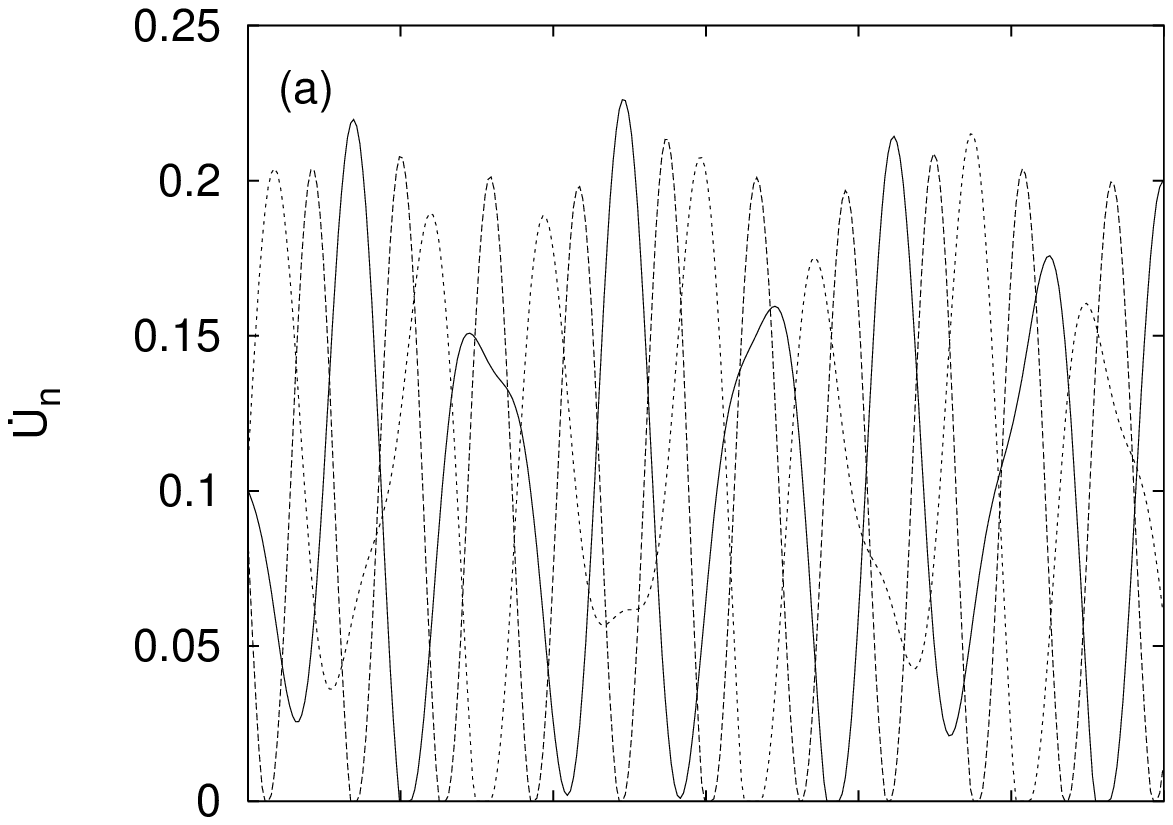}\\
\includegraphics[scale=.6]{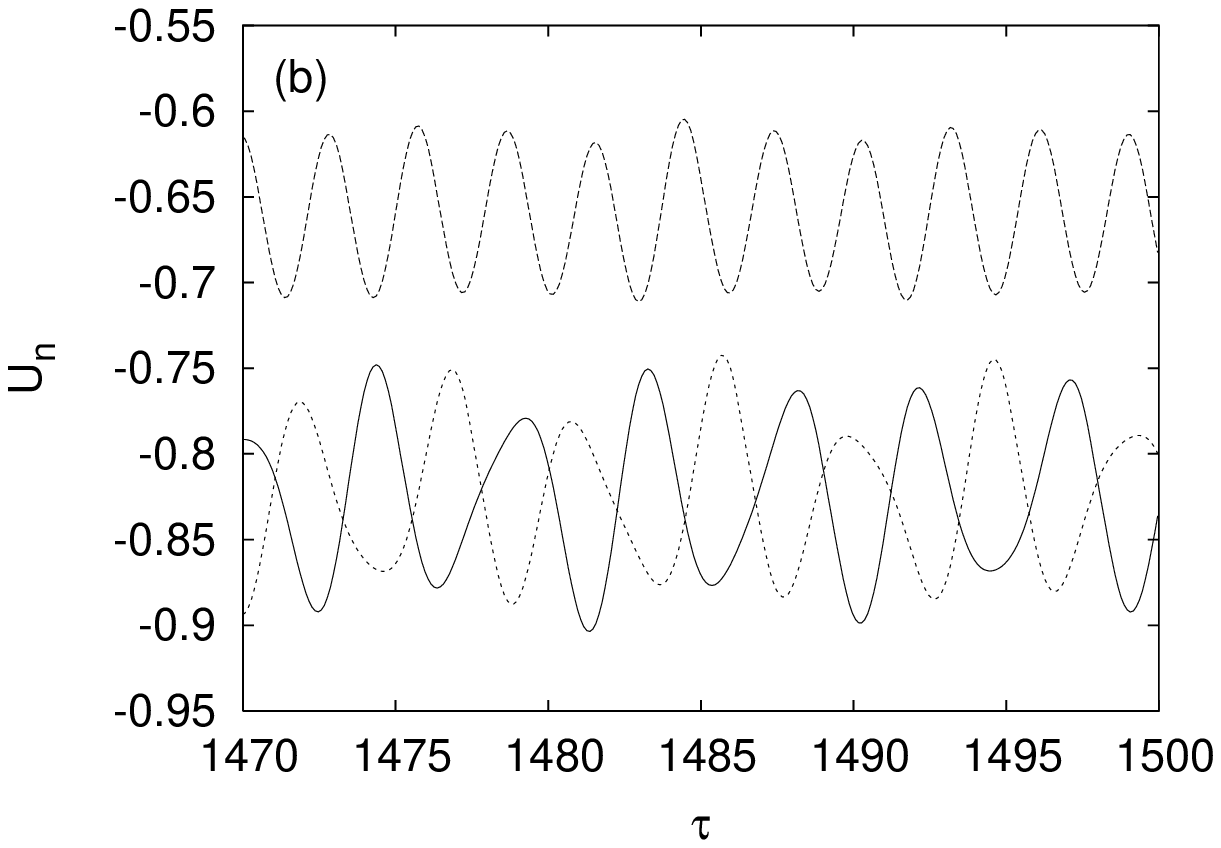}
\caption{(a) Velocities of blocks as a function of time t. (b)
 Positions of blocks as a function of time t.
 The initial position of the first block $U_1(\tau=0)=+0.989$.
 Solid line for block 1, dashed
 line for block 2 and dotted line for block 3.}
\end{figure}

\section{Conclusions}

The linearization procedure, described in Section II, does not alter the system
behaviour, if the value of the coefficient $a$ is smaller than $0.1$. This
means, that the only relevant nonlinearity in this range of $a$ is due to the stick-slip dynamics.
Nonlinear character of the velocity dependence of the friction force is not
essential for the kind of motion. This result is in accordance with our
previous conclusions on the train model of two blocks \cite{jskk}. From the
computational point of view, such a conclusion is optimistic, because the time
of calculations is much shorter for a map than for a set of differential
equations.

The history-dependent character of the motion below $a=0.1$ can be interpreted
as an example of the frequency-locking, which precedes a transition to chaos
\cite{ott}. When the initial values of the phases of the lateral blocks are
close to each other, this frequency-locking leads to the synchronization. If
they are oppose, we observe a kind of anti-synchronization, which persists in
time. Similar effect are observed also in the Hamiltonian ding-a-ling system
which can be described by coupled maps \cite{pgkk}.

A new question appears, if the frequency-locking could be observed for a
system with larger number of blocks, which imitates a real surface. This kind
of behaviour is analogous to an excitation in linear systems. What are the
consequences for the friction of macroscopic surfaces? At the beginning of
this text, we have listed various many-body nonlinear systems. Could these
excitations appear in some of this systems for larger number of subsystems? What are
the necessary conditions for this kind of motion? To answer these questions,
further research is needed, but the effort seems to be profitable.

\section*{Acknowledgments}
The simulations were carried out in ACK-CYFRONET-AGH. The
time on SGI 2800 machine is financed by the Polish Committee for Scientific
Research (KBN) with grant No. KBN/\-SGI2800/\-022/\-2002.


\end{document}